

\input harvmac

\def\mpla{{Mod.\ Phys.\ Lett.\ }{\bf A}}
\def\npb{{Nucl.\ Phys.\ }{\bf B}}

\def\plb{{Phys.\ Lett.\ }{\bf B}}
\def\prd{{Phys.\ Rev.\ }{\bf D}}

\def\zpc{Z.\ Phys.\ {\bf C}}

\def\frak#1#2{{\textstyle{{#1}\over{#2}}}}

\def\semi{;\hfil\break}
\def\threebar{{\overline{3}}}

\def\lf{16\pi^2}

\def\ktilde{\tilde k}
\def\ytilde{\tilde y}
\def\Ytilde{\tilde Y}
\def\mtilde{\tilde m}
\def\Atilde{\tilde A}
\def\Btilde{\tilde B}
\def\Ctilde{\tilde C}
\def\latilde{\tilde\lambda}
\def\abar{\bar \alpha}
\def\bbar{\bar\beta}
\def\ahat{\hat \alpha}

\def\Rhat{\hat R}
{\nopagenumbers
\line{\hfil LTH 434}
\line{\hfil hep-ph/9809250}
\line{\hfil Revised Version}
\vskip .5in
\centerline{\titlefont Infra-red stability of Yukawa and
soft-breaking  fixed points}
\vskip 1in
\centerline{\bf I.~Jack and D.R.T.~Jones}
\bigskip
\centerline{\it Dept. of Mathematical Sciences, 
University of Liverpool, Liverpool L69 3BX, U.K.}
\vskip .3in
We investigate the infra-red stability of the fixed points in the
evolution of the  Yukawa couplings,
$A$-parameters and soft scalar masses in a
broad class of supersymmetric theories. We show that the issue of stability
is essentially determined in all three cases  by the
eigenvalues of the same matrix. In a very wide range of physically
interesting theories it follows that, in the asymptotically free case,
the existence of stable infra-red
fixed points for the Yukawa couplings implies stable infra-red fixed points
for the $A$-parameters and soft scalar masses.

\Date{September 1998}}
The predictive power of the supersymmetric standard model and its 
extensions may be enhanced if the
renormalisation group (RG) running of the parameters is dominated by infra-red
(IR) stable fixed points. Typically these fixed points are for ratios; for
instance, of the Yukawa 
coupling to the gauge coupling\foot{often called 
Pendleton-Ross (PR) fixed points
\ref\Pendleton{B.~Pendleton and G.G.~Ross,
\plb 98 (1981) 291}},
 or the $A$-parameter to the 
gaugino mass. Moreover, even if these ratios have not attained their fixed 
point values at the weak scale (which in fact is usually 
the case), the couplings may be determined by quasi-fixed point behaviour
\ref\hill{C.T.~Hill, \prd 24 (1981) 691}. Here
the value of the coupling at the weak scale is 
independent of its value at the unification scale. For this 
scenario, the Yukawa couplings at the unification scale need to be
large.  There is a considerable literature devoted to the consequences 
of fixed-point and quasi-fixed point behaviour in the standard model and 
the MSSM, and in extensions 
thereof
~\Pendleton
\nref\lera{C.T.~Hill, C.N.~Leung and S.~Rao, \npb 262 (1985) 517}%
\nref\kra{P. Krawczyk and M. Olechowski, \zpc 37 (1988) 413}%
\nref\frogg{C.D.~Froggatt, R.G.~Moorhouse and I.G.~Knowles, \plb 298 (1993) 356}%
\nref\sschrmp{B. Schrempp and F. Schrempp, \plb 299 (1993) 321}%
\nref\care{M.~Carena et al, \npb 419 (1994) 213}%
\nref\bill{W.A.~Bardeen et al, \plb 320 (1994) 110}%
\nref\carwag{M.~Carena and C.E.M.~Wagner, \npb 452 (1995) 45}%
\nref\schrmp{B. Schrempp, \plb 344 (1995) 193}%
\nref\nggra{M.~Lanzagorta and G.G.~Ross, \plb 349 (1995) 319}%
\nref\fjja{P.M.~Ferreira, I.~Jack and D.R.T.~Jones,
\plb 357 (1995) 359}%
\nref\ggr{G.G.~Ross, \plb 364 (1995) 216}%
\nref\nggrb{M.~Lanzagorta and G.G.~Ross, \plb 364 (1995) 363}%
\nref\casll{A. Casas, A. Lleyda and  C. Munoz, \plb 389 (1996) 305}%
\nref\fjjb{P.M.~Ferreira, I.~Jack and D.R.T.~Jones,
\plb 392 (1997) 376}%
\nref\alamop{B.C. Allanach, G. Amelino-Camelia and O. Philipsen,
\plb 393 (1997) 349}%
\nref\bask{B.~Allanach and S.F.~King, \plb 407 (1997) 124}%
\nref\aball{S.A.~Abel and B.C.~Allanach, \plb415 (1997) 371}%
\nref\Brahm{B. Brahmachari, \mpla 12 (1997) 1969}%
\nref\aballb{S.A.~Abel and B.C.~Allanach, \plb 431 (1997) 339}%
\nref\band{M. Bando, J. Sato and K. Yoshioka, 
Prog. Theor. Phys. 98 (1997) 169}%
\nref\caesha{J.A. Casas, J.R. Espinosa and  H.E. Haber,
\npb 526 (1998) 3}%
--\ref\yejuka{G.K. Yeghian, M. Jurcisin and D.I. Kazakov, 
hep-ph/9807411}. A necessary condition for the success of both the fixed 
point and the quasi-fixed point 
scenarios is the existence of {\it stable} infra-red fixed points. Hence it is
of interest to be able to determine for a given theory the nature of the 
infra-red fixed points, and this is the question we address in this paper. 
We consider $N=1$ supersymmetric gauge theories, 
and investigate the IR stability of the fixed point structure of 
the Yukawa couplings, the soft-breaking $A$-parameters and the
soft-breaking masses\foot{Stability of the Yukawa couplings alone 
has been considered for a range of theories in Ref.~\bask}. 
The stability of the fixed points in each case is 
determined by the positivity of the eigenvalues of a matrix $M-pQI$ where 
$Q$ is the coefficient of the one-loop gauge 
$\beta$-function, and the matrix $M$
and the parameter $p$ depend on the type of coupling.
Our main result is that for a theory for which the anomalous dimension matrix 
for the matter fields is diagonal, there are matrices $R$, $S$ 
(which are $r\times s$ and $s\times r$ respectively) and $D$ (which is 
$r\times r$ and diagonal) 
such that the matrix $M$ is given by $DRS$ for the Yukawa case,
$RSD$ for the $A$-parameter case and $SDR$ for the soft mass case. It is easy 
to show that the non-zero eigenvalues of the three matrices $DRS$, $RSD$ and
$SDR$ coincide, and hence the eigenvalues determining the stability of the
fixed points for all the couplings in the theory are given in terms of one basic
set. Moreover in a more restricted class of theories (but which still includes 
all cases commonly considered) we can show that these non-zero eigenvalues are 
all positive. This implies that in the case of negative $Q$, all the 
fixed points are infra-red stable. 

For an $N=1$ supersymmetric gauge theory with gauge group $\Pi_aG_a$,
superpotential
\eqn\Ea{
W={1\over6}Y^{ijk}\Phi_i\Phi_j\Phi_k+{1\over2}\mu^{ij}\Phi_i\Phi_j, }
and soft breaking terms given by
\eqn\Eb{
L_{\rm SB}=(m^2)^j_i\phi^{i}\phi_j+
\left({1\over6}h^{ijk}\phi_i\phi_j\phi_k+{1\over2}b^{ij}\phi_i\phi_j
+ {1\over2}\sum_aM_a\lambda_a\lambda_a+{\rm h.c.}\right)}
(where $\phi$ is the scalar component of $\Phi$, and $\lambda$ is the gaugino),
the $\beta$-functions for $Y$, $h$, $b$ and $m^2$ are given by 
\eqna\Ec$$\eqalignno{
\lf\beta_Y^{(1)ijk}&=Y^{ijp}P^k{}_p+(k\leftrightarrow
i)+(k\leftrightarrow j),&\Ec a\cr
\lf\beta_h^{(1)ijk}&=h^{ijl}P^k{}_l+Y^{ijl}X^k{}_l 
+ (k\leftrightarrow i) + (k\leftrightarrow j),&\Ec b \cr
\lf\beta_b^{(1)(ij)}&=b^{il}P^k{}_l+{1\over2}Y^{ijl}Y_{lmn}b^{mn}
+\mu^{il}X^j{}_l,&\Ec c\cr
\lf[\beta_{m^2}^{(1)}]^j{}_i&=
{1\over2}Y_{ipq}Y^{pqn}(m^2)^j{}_n+{1\over2}Y^{jpq}Y_{pqn}(m^2)^n{}_i
+2Y_{ipq}Y^{jpr}(m^2)^q{}_r\cr &\quad
+h_{ipq}h^{jpq}-8M_aM_a^*g_a^2C^a_i\delta^i{}_j
+2g_a^2(R^a_A)^i{}_j\Tr[R^a_Am^2],&\Ec d\cr
}$$
where
\eqn\Ed{\eqalign{
P^i{}_j&=\frak{1}{2}Y^{ikl}Y_{jkl}-2g_a^2C^a_i\delta^i{}_j,\cr
X^i{}_j&=h^{ikl}Y_{jkl}+4Mg_a^2C^a_i\delta^i{}_j.\cr}}
Here $g_a$ is the gauge coupling for the gauge group $G_a$ and $C^a_i$ is
the quadratic Casimir for the representation $R_A^a$ of $\Phi_i$. 
The matrix $P$ is 
related to the one-loop anomalous dimension by $\gamma^{(1)}={1\over{\lf}}P$.
Now suppose that the distinct, independent couplings in the superpotential
are denoted $Y_{\alpha}$, $\alpha=1,\ldots r$, and 
$\mu_{\ahat}$, $\ahat
=1,\ldots q$, and that the distinct group 
multiplets are denoted $\Phi_I$, $I=1,\ldots s$. Let us assume that $P^i{}_j$ 
is diagonal, with $P^i{}_j=P_I\delta^i{}_j$ for all $\Phi_i$ in $\Phi_I$.
Further, let us assume that one of the couplings $g_a$ (where $G_a\ne U(1))$
and its corresponding 
gaugino mass $M_a$ are dominant in the RG evolution. Henceforth we denote 
these parameters by $g$, $M$ and suppress the remaining
gauge couplings and gaugino masses. 
We may write 
\eqn\Ee{
P_I=\sum_{\alpha}S_{I\alpha}y_{\alpha}-2g^2C_I}
where $y_{\alpha}=Y_{\alpha}^2$. Here $S_{I\alpha}$ is 
related to the dimensionality 
of the $\Phi_I$; clearly it is zero unless the superpotential term 
involving $y_{\alpha}$ contains $\Phi_I$.
The one-loop RG equation for $Y_{\alpha}$ is 
\eqn\Eea{
{d\over{dt}}Y_{\alpha}=Y_{\alpha}\sum_{I}R_{\alpha I}P_I}
where $\lf t=\ln \mu$ and $R_{\alpha I}$ takes the value 0, 1, 2
or 3 according as the superpotential term involving $y_{\alpha}$ contains 
0, 1, 2 or 3 $\Phi_I$s. 
Note that the zero entries of $R_{\alpha I}$ coincide with 
those of $S_{I\alpha}$,  and we have $\sum_IR_{\alpha I}=3$ for all $\alpha$.
As an example, for the usual MSSM superpotential, retaining only 3rd generation 
Yukawas and $g_3$,
\eqn\ssm{
W = Y_tQtH_2 + Y_bQbH_1+Y_{\tau}L\tau H_1,}
we have 
\eqn\nssma{\eqalign{
P_Q = Y_t^2 + Y_b^2 - \frak{8}{3}g_3^2,\quad P_L=Y_{\tau}^2,&\quad
P_t = 2Y_t^2 - \frak{8}{3}g_3^2,\cr\quad P_b = 2Y_b^2 - \frak{8}{3}g_3^2,\quad
P_{\tau}=2Y_{\tau}^2,\quad
P_{H_2} &= 3Y_t^2, \quad
P_{H_1} = 3Y_b^2+Y_{\tau}^2,\cr}}
so that $R_{Y_tQ} = R_{Y_bQ} =1, R_{Y_t t} = R_{Y_b b} = 1$, and 
$S_{Q Y_t} = S_{QY_b} =1, S_{t Y_t} = S_{b Y_b} = 2$ etc.

For the moment we shall focus on the parameters $Y^{ijk}$, $h^{ijk}$ and 
$m^2$; we shall discuss the $\mu$ and $b$ parameters later. 
We assume that the distinct independent 
trilinear soft-breaking couplings
are in one-to-one correspondence with the $Y_{\alpha}$, and we define 
$A_{\alpha}={h_{\alpha}\over{Y_{\alpha}}}$. We then find
\eqn\Ef{
\lf\beta_{A_{\alpha}}^{(1)}=2\sum_{I,\beta}R_{\alpha I}S_{I \beta}
y_{\beta}A_{\beta}+4\sum_{I}g^2MR_{\alpha I}C_I.}
Similarly, we find, assuming a common soft mass $m_I$ for each multiplet 
$\Phi_I$,
\eqn\Eg{
\lf\beta_{m_I^2}^{(1)}=2\sum_{\alpha,J}S_{I \alpha}y_{\alpha}
R_{\alpha J}m_J^2+2\sum_{\alpha}S_{I \alpha}
y_{\alpha}A_{\alpha}A^*_{\alpha}
-8g^2MM^*C_I.}

The one-loop RG equations for $g$ and $M$ are
\eqn\Eh{
{d\over{dt}}g=Qg^3,\qquad
{d\over{dt}}M=2Qg^2M,}
and it follows from Eqs.~\Eea, \Eh\ that the RG evolution of $\ytilde_{\alpha}
={y_{\alpha}\over g^2}$, $\Atilde_{\alpha}={A_{\alpha}\over M}$ and
$\mtilde^2_I={m^2_I\over {MM^*}}$ is given by
\eqn\Ei{\eqalign{
{d\over{dt}}\ytilde_{\alpha}&=
2\ytilde_{\alpha}\left(\sum_{I}R_{\alpha I}P_I-g^2Q\right)\cr
{d\over{dt}}\Atilde_{\alpha}&=
2g^2\sum_{I,\beta}R_{\alpha I}S_{I\beta}
\ytilde_{\beta}\Atilde_{\beta}-2Qg^2\Atilde_{\alpha}
+4\sum_{I}g^2R_{\alpha I}C_I\cr
{d\over{dt}}\mtilde^2_I&=
2g^2\sum_{\alpha,J}S_{I \alpha}\ytilde_{\alpha}
R_{\alpha J}\mtilde_J^2-4Qg^2\mtilde_I^2+2g^2
\sum_{\alpha}S_{I \alpha}
\ytilde_{\alpha}\Atilde_{\alpha}\Atilde^*_{\alpha}
-8g^2C_I.\cr}}
These equations may be rewritten in the more compact form
\eqn\Ej{\eqalign{
g^{-2}{d\over{dt}}\ytilde&=2D(RS\ytilde-U)\cr
g^{-2}{d\over{dt}}\Atilde&=2(RSD-QI)\Atilde+4RC\cr
g^{-2}{d\over{dt}}\mtilde^2&=2(SDR-2QI)\mtilde^2+2SD\Ctilde-8C,\cr}}
where $R$, $S$ and $D$ are $r\times s$, $s\times r$ and $r\times r$ matrices 
with $D_{\alpha \beta}=\ytilde_{\alpha}\delta_{\alpha \beta}$,
and the column vectors $U$ and $\Ctilde$ are defined by
\eqn\El{
U_{\alpha}=2(RC)_{\alpha}+Q, \qquad
\Ctilde_{\alpha}=\Atilde_{\alpha}\Atilde^*_{\alpha}.}
We now see the fixed point structure quite clearly. Turning firstly to the
Yukawa couplings, $\ytilde_{\alpha}=0$ is always a possible fixed point 
for each $\alpha$. The most general fixed point satisfies
\eqn\Ela{\eqalign{
\ytilde_{\alpha}&=0\quad(\alpha\in J)\cr
(RS\ytilde-U)_{\alpha}&=0,\quad(\alpha\in\{1,\ldots r\}\backslash
J)\cr}}
where $J$ is an arbitrary subset of $\{1,\ldots r\}$.
If this condition is satisfied then 
we also see that 
\eqn\Elb{\eqalign{
\Atilde_{\alpha}&=-1\quad(\alpha\in\{1,\ldots r\}\backslash
J)\cr
\Atilde_{\alpha}&={1\over Q}\left(2(RC)_{\alpha}
-\sum_{\beta}(RSD)_{\alpha\beta}\right)\quad(\alpha\in J)\cr
}} 
represents a fixed point. (If $Q=0$ and $J\ne$\O, then there are in general
no fixed points for $\Atilde_{\alpha}$ with $\alpha\in J$.)
The existence of this fixed point, corresponding  
to a universal $A$-parameter, 
$A_{\alpha} = -M$ (in the case $J=$\O), was 
first remarked in Ref.~\fjja\ (and the importance of this value 
for $A$ in the special case of a finite theory was realised earlier, in
Ref.~\ref\jmy{D.R.T.~Jones, L.~Mezincescu and Y.-P.~Yao, \plb148 (1984) 317}).
If $(SDR-2QI)$
is invertible then the fixed point for $\mtilde^2$ is given by
\eqn\Elc{
\mtilde^2=(SDR-2QI)^{-1}(4C-SD\Ctilde),}
while otherwise there is no obvious 
closed form expression. It is easy to show from Eq.~\Ej\
that in the case $J=$\O, the quantity $(R\mtilde^2)_{\alpha}$ has the fixed 
point $(R\mtilde^2)_{\alpha}=1$ for all $\alpha$. 
This ``sum rule'' has been investigated in 
Refs.~\ref\kubo{T.~Kawamura, T.~Kobayashi and J.~Kubo, \plb405 (1997) 64\semi
T.~Kobayashi et al, \npb511 (1998) 45\semi
T.~Kobayashi, J.~Kubo and G.~Zoupanos, \plb427 (1998) 291\semi
T.~Kawamura, T.~Kobayashi and J.~Kubo, \plb432 (1998) 108}.
We should mention also that in the special case 
where the Yukawa fixed points satisfy 
$(S\ytilde-2C)_I={1\over3}Q$  for all $I$
(corresponding to the so-called $P={1\over3}Q$ case\ref\jj{I.~Jack and 
D.R.T.~Jones, \plb349 (1995) 294}) 
then it is straightforward to show that $\mtilde^2_I=
{1\over3}$ is a fixed point for all $I$. 

In any event, our main concern here is with the IR stability of the fixed 
points rather than their actual values. The stability of the fixed points
of the evolution equations for a
generic set of parameters $\lambda_i$ is determined by 
linearising the evolution equations about the fixed points $\lambda_i^*$
so that they adopt
the form ${d\over{dt}}\delta\lambda_i={\partial\beta_i\over
{\partial\lambda_j}}\big|_{\lambda^*}\delta\lambda_j$, where
$\delta\lambda=\lambda-\lambda^*$. The system is 
IR stable if all the eigenvalues of ${\partial\beta_i\over
{\partial\lambda_j}}\big|_{\lambda^*}$ have positive real parts. In the case 
at hand the
task is fairly simple, the soft-breaking equations being linear already, and
the only subtlety arises for the Yukawa coupling case. We find
\eqn\Ema{
{\partial\beta_{\ytilde_{\alpha}}\over{\partial \ytilde_{\beta}}}
=2\delta_{\alpha\beta}(RS\ytilde-U)_{\beta}+(DRS)_{\alpha\beta}.}
Let us first of all consider the simplest case, of a fixed point with all
 of the
$\ytilde_{\alpha}$ non-zero, in other words $J=$\O. The Yukawa fixed points then
simply satisfy $RS\ytilde-U=0$ and it follows from 
Eq.~\Ema\ that the Yukawa 
stability matrix is just $DRS$. We then see immediately from Eq.~\Ej\ 
that the stability matrices for the
$\Atilde$ and $\mtilde^2$ are $RSD-QI$ and $SDR-2QI$. Hence for IR stability 
we require $DRS$, $RSD-QI$ and $SDR-2QI$ to have
positive eigenvalues. The appearance of combinations of $R$, $S$ and $D$ in
each of these stability matrices seems quite remarkable. Even more 
interestingly, the three matrices 
$DRS$, $RSD$ and $SDR$ have the same set of non-zero eigenvalues.
This follows from the easily-proven result that $MN$ and $NM$ have the same 
non-zero 
eigenvalues for any $m\times n$ and $n\times m$ matrices $M$, $N$. 

We have some immediate and surprisingly powerful results. 
In an asymptotically 
free theory ($Q<0$) such that the Yukawa couplings flow to non-zero PR fixed 
points, the $A$-parameters and soft masses also flow to fixed 
points. A non-asymptotically-free theory ($Q>0$) cannot have completely 
stable fixed points for both $A$-parameters and soft masses unless the 
number of Yukawa couplings and the number of fields is the same. 
This is because unless the matrices $R$ and $S$ are square, then 
either $RSD$ or $SDR$ will have some zero eigenvalues (except possibly 
in pathological cases where, $RSD$ (say) has fewer than $r$ independent 
eigenvectors).

Now let us turn to the case of a fixed point where some of the 
$\ytilde_{\alpha}$ ($p$, say) are chosen 
to be zero. Let us re-order the $\ytilde_{\alpha}$ so that $J=\{r-p+1,\ldots 
r\}$,
and let us partition the $\ytilde$ into $\ytilde_{\alpha}$, $\alpha=1,\ldots 
r-p$
and $\ytilde_{\abar}$, $\abar=r-p+1,\ldots r$. It follows from Eq.~\Ela\
that the stability matrix for the
Yukawa couplings now takes the block form 
\eqn\Emb{
\pmatrix{T_1&T_2\cr
0&T_3\cr},}
where 
\eqn\Emc{
T_{1\alpha\beta}=(DRS)_{\alpha\beta},\quad
T_{2\alpha\bbar}=(DRS)_{\alpha\bbar},\quad
T_{3\abar\bbar}=\delta_{\abar\bbar}(RS\ytilde-U)_{\bbar},}
and (with some abuse of notation) $R$, $S$ and $D$ represent the relevant 
matrices after re-ordering the couplings.
Hence for stability of the Yukawa fixed points we require 
$(RS\ytilde-U)_{\bbar}>0$ for
$\bbar=r-p+1,\ldots r$ together with the positivity of the eigenvalues of the
matrix $T_1$. The stability matrices for the $\Atilde$ and $\mtilde^2$ are
still given by $RSD-QI$ and $SDR-2QI$, and it still follows that the non-zero
eigenvalues of $DRS$, $RSD$ and $SDR$ coincide--though there are now additional
zero eigenvalues, since $D$ contains $p$ zeroes on the diagonal.
The eigenvalues of $RSD-QI$ consist of $-Q$ ($p$ times) 
together with the 
eigenvalues of $T_1-QI$. Hence this fixed point can only be stable in the 
asymptotically free case. 

Finally we should say a word about the $\mu$ and $b$ parameters. 
The evolution of the $\mu$ parameter is governed by
\eqn\emu{
g^{-2}{d\over{dt}}\mu_{\ahat}=\mu_{\ahat}(\Rhat S\ytilde-2\Rhat C)_{\ahat},}
where $\Rhat_{\ahat I}$ takes the value 0, 1 or 2 according as the $\mu$ term 
contains 0, 1 or 2 $\Phi$s. The only fixed point for $\mu$ (or any 
combination such as $\mu\over g$, $\mu\over M$) is $\mu_{\ahat}=0$
for each $\ahat$ (except in the special case of a one-loop finite theory 
with $P=0$, for which any value of $\mu$ is a fixed point). The stability 
of $\mu_{\ahat}=0$ requires $(\Rhat S\ytilde
-2\Rhat C)_{\ahat}>0$ for each $\ahat$.
In the case of a theory
with no gauge singlets, we have $Y_{lmn}b^{mn}=0$ in Eq.~\Ec{c}. In that case 
we have
\eqn\emua{
g^{-2}{d\over{dt}}\Btilde=-2Q\Btilde+2\Rhat SD\Atilde+4\Rhat C,}
where $\Btilde_{\ahat}={b_{\ahat}\over{\mu_{\ahat}M}}$. 
Eq.~\emua\ clearly determines the fixed point for $\Btilde$ in terms of that 
for $\Atilde$. Moreover, the stability of the fixed point for  
$\Btilde_{\ahat}$ simply requires that $Q$ be negative. However, in
a theory with gauge singlets, the stability of $\Btilde_{\ahat}$ is
determined by a new matrix which has no obvious connection with those which
have arisen hitherto. 

Let us illustrate our results with some simple examples. We begin with 
a simplified version of the NMSSM, with the superpotential\foot{Here 
(and also in 
the next example of the MSSM) we ignore all except the
third generation $Y_t$ coupling. It is straightforward to generalise to the case
of diagonal couplings for the other generations.}    
\eqn\nssm{
W = Y_tQtH_2 + \lambda N H_1H_2 - \frak{k}{3}N^3.}
The one-loop anomalous dimensions are given by
\eqn\nssma{\eqalign{
P_Q = Y_t^2 - \frak{8}{3}g_3^2,\quad
P_t &= 2Y_t^2 - \frak{8}{3}g_3^2,\quad
P_{H_2} = 3Y_t^2 +\lambda^2,\cr
P_{H_1} = \lambda^2,\quad P_N &= 2(\lambda^2 + k^2).\cr
}}
Writing $\Ytilde_t={Y_t \over g}$, etc, we find that 
our matrices $R, S, D$ are given in the NMSSM case by 
\eqn\nssma{
R= \pmatrix{1&1&1&0&0\cr 0&0&1&1&1\cr 0&0&0&0&3\cr}\qquad
S = \pmatrix{1&0&0\cr 2&0&0\cr 3&1&0\cr 0&1&0\cr 0&2&2\cr}\qquad
D= \pmatrix{{\Ytilde_t}^2&0&0\cr 0&{\tilde\lambda}^2&0\cr 0&0&
{\tilde k}^2\cr}.}
The Yukawa coupling evolution in this model was analysed in Ref.~\bask. 
The only non-trivial fixed point with $\Ytilde_t^2, \latilde^2,
\ktilde^2\geq 0$ 
is $$\latilde = \ktilde = 0, \quad \Ytilde_t^2 = \frak{7}{18}$$ and is 
clearly stable, since the 
stability matrix of Eq.~\Emb\ is just $\pmatrix{\frak{7}{3}&\frak{7}{18}&0\cr
0&\frak{25}{6}&0\cr 0&0&3\cr}$.   
From Eq.~\Elb\ we find the fixed point for the 
$\Atilde$ to be \eqn\nssmb{
\Atilde_{Y_t}=-1, \quad \Atilde_{\lambda}=\frak{7}{18},
\quad \Atilde_k=0,}
and from Eq.~\Elc\ we find the fixed points for the $\mtilde^2$ to be
\eqn\nssmc{
\mtilde^2_Q=\frak{41}{54},\quad \mtilde^2_t=\frak{17}{27},\quad 
\mtilde^2_{H_2}=-\frak{7}{18},}
in agreement with Ref.~\aball. Again these fixed points for the $\Atilde$ and
$\mtilde^2$ are clearly stable, since $Q$ is negative and $T_1=\frak{7}{3}$. 
Our formalism excludes the possibility of determining fixed point values
for $A$-parameters corresponding to mixing between the generations.

For our second example we take the model with the MSSM superpotential
of Eq.~\ssm. 
The matrices $R$ and $S$ are given (ordering the fields as $Q$, $L$, $t$, $b$, 
$\tau$, $H_2$, $H_1$) by
\eqn\smb{
R=\pmatrix{1&0&1&0&0&1&0 \cr1&0&0&1&0&0&1 \cr0&1&0&0&1&0&1\cr}\qquad
S=\pmatrix{1&1&0\cr 0&0&1\cr 2&0&0\cr 0&2&0\cr 0&0&2\cr 3&0&0\cr 0&3&1\cr}
\qquad D=\pmatrix{{\Ytilde}_t^2&0&0\cr 0&{\Ytilde}_b^2&0\cr 
0&0&{\Ytilde}_{\tau}^2\cr}.} 
There is no fixed point with $\Ytilde_{\tau}^2>0$.
However, in the case where we take $\Ytilde_{\tau}=0$ and $\Ytilde_t$, 
$\Ytilde_b$ non-zero
at the fixed point, we find a fixed point with 
\eqn\smc{\eqalign{
\Ytilde^2_t=\Ytilde^2_b=\frak{1}{3},&\quad \Atilde_t=\Atilde_b=-1,\quad
\Atilde_{\tau}=\frak{1}{3},\cr\mtilde^2_Q=\mtilde^2_t=\mtilde^2_b=\frak{2}
{3},\quad&\mtilde^2_{H_1}=\mtilde^2_{H_2}=-\frak{1}{3},\quad \mtilde^2_L=
\mtilde^2_{\tau}=0.\cr}} 
We find in Eq.~\Emb\ that at the fixed point 
$T_1=\pmatrix{2&\frak{1}{3}\cr\frak{1}{3}&2}$ and $T_3=1$. The non-vanishing 
eigenvalues of $T_1$ are 
$\frak{5}{3}$ and $\frak{7}{3}$. Since moreover $Q<0$ for the MSSM, 
we deduce that the fixed points are stable.  
Evidently the above fixed point values  for $Y_t, Y_b$ correspond 
to large $\tan\beta$ (specifically $\tan\beta = m_t/m_b$); 
the RG evolution of the Yukawa couplings in this region was studied 
in Ref.~\schrmp. It would be interesting to extend this analysis to
include the soft--breaking terms. 

For our third example we take a model with superpotential
\eqn\sua{
W=\sum_{i,I}^{n_g}Y_{iI}Q_it_IH_{iI}.}
We have
\eqn\sub{\eqalign{
R_{iI,i}=1, \quad R_{iI,I}&=1, \quad R_{iI,iI}=1, \cr
S_{i, iI}=1, \quad S_{I,iI}&=2, \quad S_{iI,iI}=3, \cr
D_{iI,iI}&=Y_{iI}^2/g^2=\ytilde_{iI},\cr}}
with all other entries in $R$, $S$ and $D$ being zero.
The Yukawa fixed point is $\tilde y_{iI}=\ytilde$, where 
$\ytilde={3Q+16\over{9(n_g+1)}}$, for all $i$, $I$ (assuming 
we take all the $\ytilde$ non-zero). In this case the fixed point for every
$\Atilde$ is $\Atilde_{iI}=-1$, and the fixed points for $\mtilde^2$ are
\eqn\suba{
\mtilde^2_i={1\over{3Q}}(3n_g\ytilde-8),\quad
\mtilde^2_I={1\over{3Q}}(6n_g\ytilde-8),\quad
\mtilde^2_{iI}={3\ytilde\over Q}.}
For stability it is sufficient to 
consider $RS$. 
It is straightforward to compute the eigenvalues explicitly in this 
case. We find that $RS-\lambda I$ is the $n_g^{n_g}\times n_g^{n_g}$ matrix 
given by
\eqn\suc{
RS-\lambda I=\pmatrix{[5-\lambda]I+K&2I&2I&\ldots&\ldots&2I&2I\cr
            2I&[5-\lambda] I+K&2I&\ldots&\dots&2I&2I\cr
            \ldots&\ldots&\ldots&\ldots&\ldots&\ldots&\ldots\cr
            \ldots&\ldots&\ldots&\ldots&\ldots&\ldots&\ldots\cr
            2I&2I&2I&\ldots&\ldots&2I&[5-\lambda] I+K\cr}}
where $K$ is an $n_g\times n_g$ matrix with every entry 1, and $I$ is the 
$n_g\times n_g$ unit matrix. By subtracting the last column from every other 
column, and then adding every row to the last row, we conclude that 
\eqn\sud{
\det(RS-\lambda I)=\det[(3-\lambda)I+K]^{n_g-1}\det[(3+2n_g-\lambda)I+K].}
By similar operations we can show that
\eqn\sue{
\det[xI+K]=x^{n_g-1}(x+n_g),}
and it follows that the eigenvalues of $RS$ are 3 ($[n_g-1]^2$ times),
$n_g+3$ ($[n_g-1]$ times), $2n_g+3$ ($[n_g-1]$ times) and $3n_g+3$ (once).
These are all positive. Hence again we find stability in the asymptotically
free case. It is easy to combine our second and third examples 
to obtain the $n_g$ generation theory which was considered as a model
for strong unification in Ref.~\ref\kr{S.F.~King and G.G.~Ross, 
hep-ph/9803463}.

As our final example we will take a semi-realistic GUT with gauge group  
$SU_3\otimes SU_3\otimes SU_3$, and a matter content 
consisting of $n$ sets each of the representations  $X\equiv (3, 3, 1)$,
$Y\equiv (1, \threebar ,  3)$ and $Z\equiv (\threebar , 1, \threebar )$.
The superpotential for the theory is : 
\eqn\Eq{
W = {1\over{3!}}(\lambda_1 X^3 + \lambda_2 Y^3  + \lambda_3 Z^3 ) 
+ \rho XYZ.}
Here $\lambda_1 X^3\equiv ( \lambda_1 )^{\alpha\beta\gamma} 
X_{\alpha}X_{\beta}X_{\gamma}$, where $\alpha, \beta\cdots = 1\cdots n$.  
If we set the three gauge couplings all equal to $g$ then it is easy to see 
that they remain equal under renormalisation.
In what follows we will suppose that we have 
$(\lambda_i^2)^{\alpha}_{\beta} = (\lambda_i^2)\delta^{\alpha}_{\beta}$, 
and $(\rho^2)^{\alpha}_{\beta} = \rho^2\delta^{\alpha}_{\beta}$. 
In this case we have (note that this is a case with $ s > r$)
\eqn\nssmb{
R= \pmatrix{3&0&0\cr 0&3&0\cr 0&0&3\cr 1&1&1\cr}\qquad
S = \pmatrix{2&0&0&3\cr 0&2&0&3\cr 0&0&2&3\cr}\qquad
D= \pmatrix{{\tilde\lambda}_1^2&0&0&0\cr 0&{\tilde\lambda}_2^2&0&0\cr 
0&0&{\tilde\lambda}_3^2&0\cr0&0&0&{\tilde \rho}^2\cr}.}
At the fixed point we have $6\tilde\lambda_i^2+9\tilde\rho^2=Q+16$, and
hence we must take $Q>-16$ to ensure physical fixed point values.  
We find the non-zero eigenvalues of $DRS$ etc to be $Q+16-9\tilde\rho^2$ 
(twice) and $Q+16$, which are positive provided the $\tilde\lambda_i^2$
and $\tilde\rho^2$ are positive. 
The implications for the stability of 
the infra-red fixed points were discussed in Ref.~\fjja. Again stability is 
assured in the case $Q<0$.

Each example we investigated had the property that the non-zero
eigenvalues of $DRS$ were positive (for physical fixed-point values for the 
$\ytilde$, in other words each diagonal entry of $D$ non-negative at the fixed 
point) leading to IR stability of the fixed 
points for $\Atilde$, $\mtilde^2$ for an asymptotically free theory. 
The obvious question is whether this 
is true in general. We have so far been unable to answer 
this question completely;
however, as a first step we can show that if $S$ may be written $S=D_1R^TD_2$, 
where $D_1$ 
and $D_2$ are diagonal matrices with positive entries along the diagonal, 
and if $D$ has non-negative entries, then
$DRS$ has non-negative eigenvalues. All the 
examples we have considered have this property. For instance,
in our first example, $D_1=\rm{diag}(1,2,3,3,6)$, $D_2=\rm{diag}(1,\frak{1}{3},
\frak{1}{9})$;  in our second example, $D_1=\rm{diag}(1,3,2,2,6,3,3)$, 
$D_2=\rm{diag}(1,1,\frak{1}{3})$; 
in our third example, $D_{1i,i}=1$, $D_{1I,I}=2$,
$D_{1iI,iI}=3$, $D_2=I$; and in our fourth 
example, $D_1=\frak{2}{3}I$, $D_2=\rm{diag}
(1,1,\frak{9}{2})$. It is possible to write down matrices $R$ and $S$,
subject to the constraints that $\sum_IR_{\alpha I}=3$ and that $R$ 
corresponds to a theory with $P$ diagonal, which do not have the property above;
but they do not correspond to any physically interesting model, and in any case
we have not
been able to construct any examples which lead to non-zero eigenvalues of
$DRS$ with negative real parts. It is tempting to speculate that this is a
general result.

We thus conclude that the focusing of Yukawa couplings and  soft
breakings observed for example, in Ref.~\aball\ for the MSSM  is not in
fact specific to that theory but a general  phenomenon; the existence of
stable infra-red fixed points for the Yukawa couplings implies 
(given asymptotic freedom) stable
infra-red fixed points for the $A$-parameters and soft scalar masses. In
Ref.~\fjja\ we proposed that universality of soft masses  and couplings
at the unification scale might be associated with  IR fixed points. Our
demonstration here that there exists a strong  relationship between the
IR stability of the fixed points for these masses and couplings with the
 stability of the fixed points in the Yukawa sector makes  this scenario
even more plausible.  \listrefs \bye